# Proactive TCP mechanism to improve Handover performance in Mobile Satellite and Terrestrial Networks

Preetida Vinayakray-Jani*, DAIICT, Gandhinagar, India

preeti.vinayakray@gmail.com

Sugata Sanyal, School of Technology and Computer Science,
Tata Institute of Fundamental Research, Mumbai, India

sanyals@gmail.com

*Corresponding Author

**Abstract**

Emerging standardization of *Geo Mobile Radio* (*GMR-1*) for satellite system is having strong resemblance to terrestrial GSM (Global System for Mobile communications) at the upper protocol layers and TCP (Transmission Control Protocol) is one of them. This space segment technology as well as terrestrial technology, is characterized by periodic variations in communication properties and coverage causing the termination of ongoing call as connections of Mobile Nodes (MN) alter stochastically. Although provisions are made to provide efficient communication infrastructure this hybrid space and terrestrial networks must ensure the end-to-end network performance so that MN can move seamlessly among these networks. However from connectivity point of view current TCP performance has not been engineered for mobility events in multi-radio MN. Traditionally, TCP has applied a set of congestion control algorithms (slow-start, congestion avoidance, fast retransmit, fast recovery) to probe the currently available bandwidth on the connection path. These algorithms need several round-trip times to find the correct transmission rate (i.e., congestion window), and adapt to sudden changes connectivity due to *handover*. While there are protocols to maintain the connection continuity on mobility events, such as Mobile IP (MIP) and Host Identity Protocol (HIP), TCP performance engineering has had less attention. TCP is implemented as a separate component in an operating system, and is therefore often unaware of the mobility events or the nature of multi-radios' communication. This paper aims to improve TCP communication performance in Mobile satellite and terrestrial networks.

*Keywords:* Satellite-Terrestrial Network, Inter-system handover, TCP

## 1. Introduction

Traditional terrestrial wireless or cellular networks provide mobile communication services but they remain limited to geographic coverage. High speed vehicular mobile nodes (e.g. passengers with mobile phone travelling via high speed train or car etc.) encounter the problem of network connectivity in rural or remote area where terrestrial cellular or wireless network connectivity is either limited or not feasible. Meanwhile these high speed mobile nodes impose severe constraints on cellular system in terms of Doppler Effect, handover frequency and handover execution time.

Therefore, future communication system is optimized to provide GMR (Geo Mobile Radio)/GMPRS (Geo Mobile Packet Radio Service), a satellite-optimized GSM (Global System for Mobile communication) protocol and their terrestrial counterparts, GSM/GPRS (General Packet Radio Service).These satellite systems are intended to complement and extend the existing terrestrial network to provide complete coverage [2] while getting incorporated with Internet technology for IP based data communication [3]. In fact, there are areas like tunnels, closed mines, small obstacles, where satellite connectivity suffers from non-line-of-sight events or fading. Hence when future global communication scenario offers fast and integrated services to ubiquitous users, anytime, anywhere, the seamless handover from terrestrial network to satellite or vice versa becomes essential

Nevertheless, Geo Stationary Mobile system like Geostationary Orbit Earth (GEO) satellites, will work as an important component in future data communication network. These satellites are using multiple spot antennas, where the footprint of a satellite is partitioned in small cells called spot beams. Transfer of ongoing connection to a new spotbeam or satellite is considered to be a link layer handover.

In this focused heterogeneous networking environment, the Mobile Node (MN) often encounters the instant changes in link performance and coverage [10]. However such abrupt changes in link quality or coverage either interrupts or terminates the ongoing TCP connection and eventually deteriorates the performance of the ongoing application flows, including delaying the registration process for mobility protocols [the Rendezvous (RVS) protocol for Host Identity Protocol (HIP), or binding updates to Home Agent (HA) for Mobile IP (MIP)]. Meanwhile, there is no classical TCP stack which can simultaneously handle connections efficiently on these paths with different characteristics. Meanwhile applications have different networks and TCP requirements, such as Telnet protocol values short response time (delay sensitivity), but ftp protocol values high throughput (delay tolerant). Moreover application run on phone or small PDA has limited power. So TCP should try best to avoid unnecessary retransmission to save energy.

Therefore a new Proactive TCP mechanism, aware of the flows' characteristics and ongoing changes that can adapt their behaviour to these changes, is an effective and efficient approach to solve such problem [6]. Hence here the attempt is to make the seamless handover more efficient by proactively adapting the TCP transmission windows to adapt the network environment so that congestion and buffer overflow situations can be prevented effectively and Mobile Nodes (MN) can experience the seamless mobility with consistent quality.

However there are several approaches to deal with disruption in connectivity issue like Delay/disruption tolerant Networking [17], Proxy enhancing Proxy (PEP) [16], Accelerated TCP [15], or combination of these approaches.

The proposed proactive handover engine enables the MN to perform inter-system handover to maintain the seamless connectivity, where connection to target network is made before breaking the current one.

## 2. Background

There has been past work to adjust TCP's receive window to suppress the TCP sender from sending further data, and to adjust the receive window to be better suitable for slow modem link. However, this work has targeted environments with static and fixed setup, and has not been applied in a dynamic network environment with different types of applications and multiple available networks with terminal mobility [8]. Moreover this work is targeted for mobile environment where TCP window size is dynamically evaluated and earlier work does not cover the problem that has been proposed here.

[1] proposes a handover prioritization scheme for different channel allocation techniques in satellite network, where handover requests are queued for predefined time interval. If there are no resources available during this interval then ongoing connectivity of services will get terminated. Also [4] proposes time based channel reservation by reserving channel in next targeting spotbeam to improve handover performance.

There has been few research approaches on adjusting the TCP's slow-start threshold based on explicit indication of available bandwidth on the path. This requires modifications to the TCP sender side that is often a fixed server and therefore inaccessible to the mobile terminal vendors [9]. This work is focused on TCP receiver that is in most cases the mobile terminal. Therefore the earlier work does not overlap with the solution this paper suggests.

There is a paper to set the receiver window to zero during the handoff to suspend the sender and avoid packet losses [11]. Here the approach was to set the receiver window directly based on bandwidth-delay product. There is some past work on receive buffer auto tuning based on bandwidth-delay product, but this does not discuss the vertical handoff [12]. This proposed work requires TCP modification on receiver side where as the sender side is required to enter in slow start phase following handover. In [14] the proposed work suggests that buffer of all links should be configured with maximum bandwidth delay product. However such work requires reliance on operator and this is somewhat difficult to justify when there is a link with low bandwidth-delay product, this may cause excess queuing.

The inter segment handover proposed in [5] considers the different phases of handover like initiation, decision and execution for different handover controlling schemes, based on where the handover initiation or decision phases are executed, in Mobile node or in network.

The work proposed in [18] closes the *congestion window* (*cwnd*) at the sender before the inter-system handover execution takes place. When inter-system handover re-establishes the connection in new TCP segment, *cwnd* is reset to 1 and Slow-Start Threshold (*ssthresh*) is set equal to estimated bandwidth for the new link. But this method does not seem to provide seamless handover from Wi-Fi to satellite.

## 3. Reference architecture: Network Connectivity

Reference architecture is shown in Figure 1 where multi-mode Mobile Node (MN) encounters the connectivity signals from three different networks when on move.

While being on vehicle it receives connectivity from 3G terrestrial network in urban area. Once vehicle approaches a tunnel, Wi-Fi connection is used by MN by means of inter-system handover from 3G to Wi-Fi. Once the vehicle exits from tunnel and enters in rural area, MN gets connected through geostationary satellite. MN receives the IP address from corresponding network while getting connected.

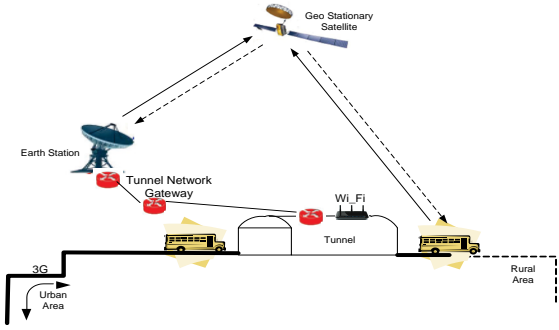

**Figure 1 Reference Architecture: Network Connectivity**

The mobility and inter-system handover events need to be managed at both Layer 2 (L2) and Layer 3 (L3) for the selection of new network link and for configuring system with new IP address. Limiting the scope of solution to L3 perspective work adopts the Mobile Internet Protocol Version 6 (MIPv6). It accompanies an efficient rerouting feature. However the proposed solution is also applicable to MIPv4. These adopted mobility protocol requires the registration of received IP addresses in corresponding Home Agent (HA), so MN can maintain end to end connectivity while on the move.

Alternatively Proxy Mobile IP (P-MIP) is located within the immediate gateway/router of terrestrial cellular networks' Service Access Point (SAP) in base station or Wi-Fi network's SAP in access router/gateway or satellite earth station interface with terrestrial GMR1. Interface with SAP is needed to reroute the Binding Update (BU) towards HA for corresponding MN. Thus if MN is not optimized with MIPv6/MIPv4, it is able to maintain end to end connectivity, through P-MIP.

However the performance of TCP is very much affected by periodic variation of radios or high mobility of MN, makes connection of MN to any network alter stochastically.

There are few parameters which needs to be considered during handover phase as they play important role for TCP performance. Such parameters include duration of handover phase, the coverage area overlapped with two different networks, speed of vehicle and buffer size available at sender, receiver and network. Aim of this paper is to provide TCP performance improvement in two possible inter-system handover cases in mobile node, from terrestrial network to satellite and from satellite to terrestrial network.

### 3.1. WLAN / GPRS → Satellite

When MN detects the imminent possibility for handover for currently active flow(s) it can disseminate the window size to peer or Corresponding Node (CN) and that can be set as:

$$W_{Default} > SATWin_{max} \geq W_{REC} \quad (1)$$

When previously connected to the Satellite network, the MN can try to estimate the maximum window size offered by satellite through bandwidth-delay product of the wireless link. MN also has the default value ($W_{Default}$) of the maximum window size that is usually larger than the bandwidth-delay product of the path. Hence it is quite obvious that in a network such as Satellite or GPRS the maximum bandwidth delay product offered is considered to be less than the default value used by TCP. Hence MN advertises a smaller window ($W_{REC}$) size to the CN, i.e. $SATWin_{max}$. If such value is larger than the default value, the TCP sender can flood the cellular network that can cause excess queuing at 2G/3G base station rather than buffer overflow.

Alternatively one can assume that MN has some previously estimated value of Satellite bandwidth delay product. Using this cached information MN selects the value for the receive window size and advertises it to the CN in a TCP acknowledgement (ACK) packet at $t_{a0}$, as shown in figure 2. Hence MN will deliver new window size before initiating its registration of a new location address to the RVS or HA. Upon receiving the window advertisement for future flows at $t_{a1}$ CN limits its data transmission rate as per the new received window size. Thus new TCP window adjustment at the CN actively reduces the network congestion by controlling the rate at which the CN communicates. Subsequently MN delays its registration process of received new location for a period of δ and sends the registration request at $t_{r0}$.

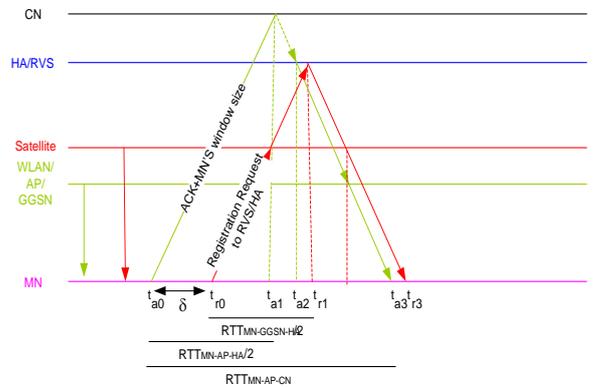

**Figure 2 Operational View of Inter-system Handover**

After registration at RVS/HA at $t_{r1}$, RVS/HA redirects all the packets to new located address. The last TCP packet sent just before the use of the new window size encountered by the CN is received by RVS/HA at time $t_{a2}$ and forwarded to the WLAN link. Therefore if $t_{r1} \ll t_{a2}$ then some packets are redirected to the GPRS network, before the influence of new window takes place. This will create buffer queue to be filled to overflowing. To avoid this problem, the value of δ should be chosen so that $t_{r1} \gg t_{a2}$. Hence MN should determine the value of δ. It can be seen from the figure 2,

$$\delta \leq RTT_{MN\text{-}SAT\text{-}CN} - 0.5(RTT_{MN\text{-}SAT\text{-}RVS/HA} + RTT_{MN\text{-}GGSN\text{-}RVS/HA}) \quad (2)$$

Once the possibility of handover is detected, then δ interval handover decision can be made and inter-system handover execution phase starts. As a result then either MN or P-MIP interface in GMR1 interface of satellite generates registration request through BU towards HA. The MN or P-MIP will receive BU ack at $t_{r3}$. It is important to note that HA is capable to register more than one binding update requests for given MN. Thus if MN remains connected to geostationary satellite network then seamless beam handover or intra-system handover requires only L2 handover as IP address received by MN from satellite network remains same.

Thus similar type of analogy can be applied to intra-system handover also.

### 3.2. Satellite → WLAN/GPRS

When handover from Satellite to GPRS/WLAN or any terrestrial network occurs, new packets reach the TCP receiver through the GPRS/WLAN link while some of the older packets are still in transit over the satellite network. Because packets arrive at the receiver in wrong order, the receiver generates duplicate acknowledgements that cause a retransmission at the TCP sender. This retransmission is unnecessary, because no packets have been lost and it simply degrades TCP throughput. Therefore, prior to handover from satellite to terrestrial network, it would be useful to temporarily close the TCP's advertised window for a short while until all outstanding packets from satellite link have arrived, to prevent packet reordering from happening.

However reordering approach defined in [19] is a sender and receiver side modification that allows managing the arrival of out-of-sequence packets due to the intersystem handover from high RTT (Round Trip Time) satellite network to low RTT WLAN (WLAN). Instead of DupACKS (Duplicate Acknowledgements), ACKs are sent and a congestion avoidance phase is performed during the interval for which out-of-sequence packets are received

The handover from satellite to GPRS/WLAN encounters the problem of under-utilization; as packet processing in satellite may take longer time due to the absence of sufficient duplicate ACKs to trigger the fast transmission of the packets over the faster GPRS/WLAN. Hence even though the advertised window size is larger, the packet remains enqueued for a long time in satellite network due to lack of duplicate ACK packets from GPRS/WLAN. Smaller bandwidth (compared to satellite) of GPRS/WLAN capacity can be underutilized because the satellite link may use significantly smaller size of advertised window than what the bandwidth-delay product of GPRS/WLAN would require. However there are attempts made by NASA to make use of TCP with Large Window (LW) – TCP-LW.

To handle the packet loss while performing the handover to GPRS/WLAN, MN increases the current satellite window size with bandwidth-delay product. The window should be set appropriately to avoid fragmentation. And this will permit CN to send more data benefiting from the fact that a CN can continue to increase its congestion window size when the amount maximum threshold of received window size is crossed. This may create the burst of data packets as well as duplicate ACKs that can trigger the fast transmission of the packets queued at Satellite. Increasing the advertised window size more than two segments may make CN to overflow the network. Once handover is performed the gradual increase in advertised receiver window size makes consistent utilization of WLAN network.

### Conclusion

The TCP window advertisement permits end-systems to actively reduce the congestion (in any TCP queue) by controlling the rate at which end-systems communicate. While performing handover from fast network to slow network, the network heads toward the congestion, the TCP flows are rate shaped by adjusting the receivers' advertised window size and simultaneously slowing down the return of the ACKs, as end-systems communications align toward the consistent possible rate. Meanwhile, when MN is supported with multi-homing mobility feature, it is very likely that such advanced mobility protocols are able to move simultaneously running traffic flows, independently from one access interface to another. Hence simultaneous flows, relevant to different or single applications may have different throughput or bandwidth requirements. Such requirements are directly influenced by the received window size parameter *TcpWindowSize*. Therefore proposed proactive handover engine allocates the proportional TCP window size by considering the type of the access network availability, single flow requirement and available buffer capacity. In proposed innovative work the MN advertises Maximum *TcpWindowSize* to CN along the ACK at appropriate time before initiating for handover and it will follow such analogy for subsequent TCP packets also. Hence proposed method is applicable when both end-systems are mobile. The proposed method is versatile as it can operate over any link layer technology e.g. Wi-MAX, WCDMA, Satellite etc. and hence it supports the Media Independent handover standardization activity, as well. The proposed method does not require any modification or enhancement in access networks.

However, further study is needed to compare the proposed work with the NASA Space Communications Protocol Specification-Transport protocol (SCPS-TP), including transport layer alternative algorithm proposed in [13]. As satellite is an open space global communication, TCP has to maintain its reliability property even when network is prone to packet loss, rendering and duplication. Besides, the TCP performance improvement also requires the MN to measure the round trip time more accurately. Hence it is desirable to implement reliability with an optimum flow control; this gives scope for Steganographic Coding [20]. As a part of further enhancement of proposed solution, various TCP header fields will be examined so that they can be further strengthened with application of Steganography, to maintain the confidentiality of TCP signalling and data.